\documentclass[letterpaper,titlepage,11pt]{article}
\usepackage{latexsym,amssymb,amstext,amsmath}
\usepackage{slashed}
\usepackage{mathrsfs}
\usepackage{color}
\usepackage{enumerate}
\usepackage{graphicx}
\usepackage{tikz}
\usepackage{etex}
\usepackage{latexsym}
\usepackage{cite}
\usepackage{caption}
\allowdisplaybreaks

\usepackage[
colorlinks=true,
linkcolor=blue,
urlcolor=red,
filecolor=green,
citecolor=red,
pdfstartview=FitV,
pdftitle={},
pdfauthor={Mehmet Ozkan},
pdfsubject={},
pdfkeywords={},
pdfpagemode=None,
bookmarksopen=true
]{hyperref}

\newcommand{\bea}{\setlength\arraycolsep{2pt} \begin{eqnarray}}
	\newcommand{\eea}{\end{eqnarray}}
\newcommand{\nn}{\nonumber}

\newcommand{\opsi}{\bar{\psi}}

\usepackage{hyperref}

\def\rmi{{\rm i}}
\newsavebox{\uuunit}
\sbox{\uuunit}
{\setlength{\unitlength}{0.825em}
	\begin{picture}(0.6,0.7)
		\thinlines
		\put(0,0){\line(1,0){0.5}}
		\put(0.15,0){\line(0,1){0.7}}
		\put(0.35,0){\line(0,1){0.8}}
		\multiput(0.3,0.8)(-0.04,-0.02){12}{\rule{0.5pt}{0.5pt}}
		\end {picture}}

	\def\be{\begin{equation}}
		\def\ee{\end{equation}}
	\def\ba{\begin{array}}
		\def\ea{\end{array}}
	\def\bea{\begin{eqnarray}}
		\def\eea{\end{eqnarray}}
	\def\bd{\begin{displaymath}}
		\def\ed{\end{displaymath}}
	
	\def\nn{\nonumber}
	
	\def\a{\alpha}
	
	\def\g{\gamma}
	\def\G{\Gamma}
	\def\d{\delta}
	
	\def\e{\epsilon}
	\def\ve{\varepsilon}
	
	\def\vf{\varphi}
	
	\def\p{\psi}

	\def\l{\lambda}
	\def\L{\Lambda}
	\def\m{\mu}
	\def\n{\nu}
	\def\r{\rho}
	\def\s{\sigma}
	
	\def\t{\tau}

	\def\o{\omega}
	\def\O{\Omega}

	\def\nn{\nonumber}
	\def\cM{\mathcal{M}}
	\def\cN{\mathcal{N}}
	\def\cV{\mathcal{V}}
	\def\cL{\mathcal{L}}
	\def\cF{\mathcal{F}}
	\def\cD{\mathcal{D}}

	\makeatletter
	\@addtoreset{equation}{section}
	\makeatother
	
\begin{document}
		%
		\begin{titlepage}
			
			\bigskip
			
			\begin{center}
				{\LARGE\bfseries Off-Shell $\cN=(1,0)$ Linear Multiplets in Six Dimensions}
				\\[10mm]
				\textbf{Ugur Atli, Omer Guleryuz and Mehmet Ozkan}\\[5mm]
				\vskip 25pt

				{\em  \hskip -.1truecm Department of Physics, Istanbul Technical University,  \\
					Maslak 34469 Istanbul, Turkey  \vskip 5pt }
				
				{email: {\tt atliugu@itu.edu.tr, omerguleryuz@itu.edu.tr,  ozkanmehm@itu.edu.tr}}
				
			\end{center}
			
			\vspace{3ex}

			\begin{center}
				{\bfseries Abstract}
			\end{center}
			\begin{quotation} \noindent
				
				We provide a tensor calculus for $n$-number of $\cN = (1,0)$ linear multiplets in six dimensions. The coupling of linear multiplets is encoded in a function $\cF_{IJ}$ that is subject to certain constraints. We provide various rigid and local supersymmetric models depending on the choice of the function $\cF_{IJ}$ and provide an interesting off-diagonal superinvariant, which leads to an $R^2$ supergravity upon elimination of auxiliary fields.

			\end{quotation}
			
			\vfill
			
			\flushleft{\today}
		\end{titlepage}
		\setcounter{page}{1}
		\tableofcontents
		
		\newpage

		\section{Introduction}{\label{Intro}}
		
		Six dimensional supergravity theories \cite{6DRef1,6DRef2,6DRef3,6DRef4,6DRef5,6DRef6,6DRef7} have been studied from various perspectives as they can be helpful in our understanding of the fundamental properties of the nature. The $\cN = (1,0)$ gauged theory, known as the Salam - Sezgin model \cite{SalamSezgin}, spontaneously compactifies on $M_4 \times S^2$ with $\cN = 1$ supersymmetry, thus becoming a natural starting point of  phenomenological studies with a string theory origin \cite{SalamSezginString}. When extended with certain higher derivative terms, six dimensional $\cN = (1,0)$ supergravity provides a useful testbed for checking proposals regarding string duality at higher orders \cite{MinasianLiu1,6DGB,MinasianLiu2}. Six dimensional supergravity theories also found themselves applications in $AdS_3/CFT_2$ correspondence \cite{deBoer}, thus providing a framework for studying the superconformal field theory in two dimensions.
		
		Initial works on six-dimensional supergravity were based on the Noether procedure and the construction of $\cN = (1,0)$ supergravity as well as its matter couplings was put into a systematic approach using superconformal tensor calculus in \cite{Bergshoeff6D}. This approach, which was originally developed in \cite{SCTC1,SCTC2,SCTC3,SCTC4}, is an off-shell methodology based on enhancing the symmetries as much as possible which restricts the possible couplings of fields within a multiplet and ease the construction of an action principle. In six dimensions, the conformal $\cN = (1,0)$ supergravity is based on the superalgebra $OSp(6,2|1)$ with the generators
		\bea
		P_a\,, M_{ab}\,, D\,, K_a \,, U_{ij} \,, Q_\alpha^i\,, S_\alpha^i \,,
		\eea
		where $a,b,\ldots $ are the Lorentz indices, $\a$ is a spinor index and $i,j = 1, 2$ is an $SU(2)$ index. Here, the first four generators, $\{P_a, M_{ab}, D, K_a \}$ are the generators of conformal algebra. $U_{ij}$ is the generator of $SU(2)$ R-symmetry and $Q_\alpha^i$ and $S_\alpha^i$ are the generators of the $Q-$SUSY and the $S-$SUSY respectively. In the superconformal approach to supergravity, one associates a gauge field to each generator and imposes a set of constraints to relate the gauge theory of $OSp(6,2|1)$ superalgebra to gravity. For six-dimensional $\cN = (1,0)$ theory, such a set of constraints separates the gauge fields in a dependent and independent set of fields such that the independent set does not provide an equal number of off-shell bosonic and fermionic degrees of freedom. This can be cured by adding a matter content, which is not unique and lead to two different Weyl multiplets: the standard Weyl multiplet and the dilaton Weyl multiplet \cite{Bergshoeff6D}. It is, however, not possible to obtain a consistent two-derivative supergravity with a standard Weyl multiplet\footnote{A consistent model for the standard Weyl multiplet can be achieved by the addition of higher curvature terms, see \cite{6DHD}.} but a dilaton Weyl multiplet coupled to a scalar (hyper) \cite{Bergshoeff6D,Sokatchev}, or a non-linear \cite{Bergshoeff6D}, or a real $O(2n)$ \cite{O2nMultiplet} or a linear multiplet is necessary \cite{Bergshoeff6D}. If an off-shell supergravity is required, then the linear multiplet is the simplest and the most utilized option whose gauge fixing of redundant superconformal symmetries lead to an off-shell six dimensional $\cN=(1,0)$ supergravity.  
		
		In \cite{6DHD,OzkanThesis}, it was shown that the linear multiplets are also essential in the construction of higher derivative models. Thus, following the previous work on four dimensional tensor multiplets \cite{deWit4D} and five dimensional linear multiplets \cite{Ozkan5D}, our aim in this paper is to provide a detailed investigation of six dimensional $\cN = (1,0)$ linear multiplets, their rigid and supergravity couplings and their higher derivative actions. As we will discuss in the next sections, the couplings of $n$-number of linear multiplets are controlled by a function $\cF_{IJ}$ which is a function of $SU(2)$ triplet of scalars $L_{ij}$ of the linear multiplet. The function is not completely free but constrained by symmetries of the theory, although the restrictions are quite generic and mild and we provide various possible choices of $\cF_{IJ}$. 
		
		This paper is organized as follows. In Section \ref{Sec2}, we first introduce rigid vector and linear multiplet and their supersymmetric coupling. The vector multiplet is an essential element in the construction of an action principle for the linear multiplets and their coupling gives rise to an a rigid linear multiplet action once we express the elements of the vector multiplet in terms of a proper combination of the elements of the linear multiplet. At this step, we introduce the function $\cF_{IJ}$ which completely determines the interaction of linear multiplets. Various choices of $\cF_{IJ}$ as well as rigid higher derivative models are discussed. Finally, we provide the dimensional reduction of vector and linear multiplets to 5D. In Section \ref{Sec3}, we introduce local superconformal two-derivative vector and linear multiplet actions. The linear multiplet action can be obtained both in the standard and the Weyl multiplet background. For the vector multiplet, the necessity for a compensating scalar for scaling symmetry implies either the use of dilaton Weyl multiplet or a linear multiplet. From a more minimalist approach, we provide a model in a dilaton Weyl multiplet. Nonetheless, we discussed an ansatz for a conformal vector-linear multiplet action, which we plan to address its details in a future publication. Next, in Section \ref{Sec4} we gauge fix and provide various supergravity models. These models include off-shell Poincar\'e supergravity both in Einstein and Jordan frames, the supersymmetric coupling of linear multiplets to supergravity and an off-diagonal $R Y_{ij}$ invariant which would lead to an on-shell six dimensional $R^2$ supergravity upon coupling to an off-shell vector multiplet and the elimination of auxiliary fields. We give our comments and conclusions in Section \ref{Sec5}.


		\section{Rigid Linear Multiplet Couplings}{\label{Sec2}}
		
		The six dimensional ${\mathcal{N}}=(1,0)$ linear multiplets can be realized off-shell in a general superconformal background. In this section, we  focus on its rigid supersymmetric realization on flat Minkowski background. The linear multiplet consists of an $SU(2)$ triplet of scalars $L_{ij}$, a tensor gauge field $E_{\mu\nu}$ and an $SU(2)$ Majorana spinor $\varphi^i$ of negative chirality. The supersymmetry transformation rules $(\epsilon^i)$ can be given as follows\footnote{We use the conventions of \cite{Coomans6D}. In particular, the spacetime signature is $(-, +, +, +, +,+)$ and $\p^i \g_{(n)} \chi^j = t_n \chi^j \g_{(n)} \p^i$ where $t_0 = t_3 = t_4 = -1$ and $t_1 = t_2 = h_5 = t_6 = 1$. When $SU(2)$ indices on spinors are omitted, northwest -southeast contraction is understood.} \cite{Bergshoeff6D,Coomans6D}
		\begin{eqnarray}
			\delta L^{ij} &=& \bar{\epsilon}^{(i} \varphi^{j)}\nonumber \,,\\
			\delta \varphi^i &=& \frac12 \slashed{\partial} L^{ij} \epsilon_j - \frac14 \slashed{E} \epsilon^i \nonumber\,,\\
			\delta E_{\mu\nu} &=& \bar\epsilon \gamma_{\mu\nu} \varphi \,,
			\label{LinearRigidSUSY}
		\end{eqnarray}
		where the $Q$-supersymmetry parameter, $\epsilon^i$, is of positive chirality. Furthermore, $E_\mu$ is a constrained vector, $\partial^\mu E_\mu = 0$, and is related to the tensor gauge field $E_{\mu\nu}$ via 
		\begin{eqnarray}
			E^\mu = \partial_\nu E^{\mu\nu} \,.
		\end{eqnarray}
		
		\subsection{Composite Vector Multiplets}{\label{SSec21}}
		
		Supersymmetric two-derivative actions for linear multiplets can be constructed by using vector multiplets as composite multiplets. The vector multiplet consists of a vector field $W_\mu$, an $SU(2)$ Majorana spinor of positive chirality $\Omega^i$ and a triplet of scalar fields $Y^{ij}$. The supersymmetry transformations are given as \cite{Bergshoeff6D, Coomans6D}
		\begin{eqnarray}
			\delta W_\mu &=& - \bar{\epsilon} \gamma_\mu \Omega \,,\nn\\
			\delta \Omega^i &=& \frac{1}{8} \gamma \cdot F \epsilon^i - \frac12 Y^{ij} \epsilon_j  \,,\nn\\
			\delta Y^{ij} &=& - \bar{\epsilon}^{(i} \slashed\partial \Omega^{j)}  \,,
			\label{VectorRigidSUSY}
		\end{eqnarray}
		where $F_{\mu\nu} = \partial_\mu W_\nu - \partial_\nu W_\mu$. The key point in the construction of linear multiplet actions is that the supersymmetric coupling of a linear multiplet to a vector multiplet takes the following form \cite{Bergshoeff6D, Coomans6D}
		\begin{eqnarray}
			{\mathcal L} &=& Y_{ij} L^{ij} + 2 \bar\Omega \varphi + \frac14 F_{\mu\nu} E^{\mu\nu} \,.
			\label{VectorLinearAction}
		\end{eqnarray}
		As was already demonstrated in four \cite{deWit4D} and five dimensions \cite{Ozkan5D}, such auxiliary expressions can be used to derive supersymmetric Lagrangians for  $n$-linear (or vector) multiplets. Such a construction is based on the observation that it is possible to construct composite expressions from the elements of the linear (or vector) multiplet that precisely transform as a vector (or linear) multiplet. Once such an expression is obtained, it can be used in the auxiliary action (\ref{VectorLinearAction}) to construct a supersymmetric linear (or vector) multiplet action.
		
		To construct an interacting $n$-linear multiplets, we follow the footsteps of \cite{deWit4D, Ozkan5D} and introduce a real function ${\mathcal F}_{IJ}(L)$ which is a function of the linear multiplet scalars $L_{ij}$. Note that the index $I,J=1,2,\ldots,n$ label the number of linear multiplets. A reasonable guess would be to start the construction by setting the lowest element of the vector multiplet as $\Omega^{i}_I = \mathcal{F}_{IJ} \slashed\partial \varphi^{iI}$, however, if $\cF_{IJ} \neq \d_{IJ}$, then such a starting point does not have the transformation structure given by (\ref{VectorRigidSUSY}). A starting point that gives rise to the correct structure is 
		\bea
		\Omega^i_I &=& {\mathcal{F}}_{IJ}\slashed \partial \varphi^{iJ} + \cF_{IJKjk} \slashed\partial L^{ijJ} \varphi^{kK} - \frac12 \cF_{IJK}{}{}^{ij} \slashed{E}^J \varphi^{K}_j + \frac16 \cF_{IJKL}{}^{ijkl} \g^a \vf_j^J \bar\vf_k^K \g_a \vf_l^L  \,,
		\label{CompositeOmega}
		\eea
		where
		\bea
		\cF_{IJKij} = \frac{\partial \cF_{IJ}}{\partial L^{ij K}} \,, \qquad \text{and} \qquad 	\cF_{IJKLijkl} = \frac{\partial \cF_{IJ}}{\partial L^{ij J} \partial L^{kl K}} 
		\eea
		The necessity that composite expression for $\Omega_I^i$ transforms like the fermionic component of the vector multiplet implies three important constraints on $\cF_{IJ}$ and its descendants
		\bea
		\cF_{IJK}{}^{ij} = \cF_{I(JK)}{}^{ij} \,, \qquad \cF_{IJKi[jk]l} = 0 \,.
		\label{RigidConstraints}
		\eea
		Upon varying the composite expression (\ref{CompositeOmega}) and insisting on the transformation rules (\ref{VectorRigidSUSY}), we obtain the following composite expressions for $Y_{ij}^I$ and $F_{ab}^I$
		\begin{eqnarray}
			Y^{ij}_{I}&=& - \cF_{IJ} \Box L^{ijJ} - \cF_{IJKkl} \partial_a L^{ik J} \partial^a L^{jl K} - \cF_{IJKk}{}^{(i} \partial_a L^{j)k J} E^{a K} - \frac14 \cF_{IJK}{}^{ij} E_a^J E^{a K} \,,\nn\\
			&& - \frac14 \cF_{IJK}{}^{ij} \bar\varphi^J \slashed\partial \varphi^K - 2 \cF_{IJK}{}^{k(i} \bar\vf_k^J \slashed \partial \varphi^{j) K} -  \cF_{IJKL}{}^{pqk(i} \partial_a L^{J j)}{}_k \bar\varphi_p^K \gamma^a \varphi_q^L \nn\\
			&& + \frac12  \cF_{IJKL}{}^{k(ij)l}  \bar\vf^J_k \g^a \varphi_l^K  E_a^L - \frac{1}{12} \cF_{IJKLM}{}^{ijklmn}\bar\vf_k^J \g_a \vf_l^K \bar\vf_m^L \g^a \vf_n^M \,,\nn\\ 
			F_{ab I}&=&  - 2 \partial_{[a} \left(  \cF_{IJ} E_{b]}^J  \right) + 2 \cF_{IJKj}{}^k \partial_{[a} L^{jl J} \partial_{b]} L_{lk}^K  + 2 \partial_{[a} \left(\cF_{IJK}{}^{ij} \bar\vf_i^J \g_{b]} \varphi_j^K\right)\,.
			\label{RigidYandF}
		\end{eqnarray}
		Before we proceed to the actual construction of the action, a couple of remarks on these composite formulae are in order. First of all, these composite expressions match with the ones in \cite{Bergshoeff6D,Coomans6D} if only a single linear multiplet is chosen. In term of $\cF_{IJ}$ such a choice would correspond to setting $\cF_{11} = L^{-1}$. In the same spirit, an $n$-number of non-interacting linear multiplets can be obtained by setting $\cF_{IJ} = \delta_{IJ} \left(L^I\right)^{-1}$ where $(L^I )^2=  L_{ij}^I L^{ij I}$ \cite{deWit4D}. Second, we note that although $F_{ab}$ is closed, i.e. $\partial_{[a}F_{bc]}^I = 0$, it is not exact \cite{deWit4D}. Consequently, the present form of  the vector-linear action (\ref{VectorLinearAction}) is essential to us although its last term can be written as $A_\mu E^\mu$ by means of an integration by parts. Finally, it is important to note that the index $I$ is fixed while the index $J$ is being summed over. This is also reflected by the fact that $\cF_{IJ}$ has no particular symmetry in the $(IJ)$ indices. This was first noted in \cite{Ozkan5D} and it was shown that a non-symmetric choice plays an important role in the construction of higher derivative superinvariants. 
		
		It is also possible to construct vector multiplet actions using the auxiliary action (\ref{VectorLinearAction}) as well as a composite linear multiplet that is constructed from the elements of the vector multiplet.
		\bea
		L_{ij} = Y_{ij}\,, \qquad \vf^i = - \slashed\partial \Omega^i \,,\qquad E^a = \partial_b F^{ba} \,.
		\label{RigidVectorMap}
		\eea
		The vector multiplet action will be of particular use in constructing higher derivative actions for linear multiplets.

		\subsection{Supersymmetric Linear and Vector Multiplet Actions}{\label{SSec22}}
		
		With the composite fields in hand, we now proceed to give the rigid supersymmetric linear multiplet action. Substituting the composite formulae (\ref{CompositeOmega}) and (\ref{RigidYandF}) into the auxiliary action (\ref{VectorLinearAction}) we obtain
		\bea
		\cL_L &=& - \cF_{IJ} L_{ij}^I \Box L^{ijJ} - \cF_{IJKkl} L_{ij}^I  \partial_a L^{ik J} \partial^a L^{jl K} - \cF_{IJKk}{}^{i}  L_{ij}^I  \partial_a L^{jk J} E^{a K} -  \frac12  \cF_{IJ} E^{aI} E_{a}^J \nn\\
		&& - \frac14 \cF_{IJK}{}^{ij}  L_{ij}^I  E_a^J E^{a K}    + \frac12 \cF_{IJKj}{}^k E^{ab I}  \partial_{a} L^{jl J} \partial_{b} L_{lk}^K - \frac14 \cF_{IJK}{}^{ij} L_{ij}^I \bar\varphi^J \slashed\partial \varphi^K   \nn\\
		&&  - 2 \cF_{IJK}{}^{ki} L_{ij}^I  \bar\vf_k^J \slashed \partial \varphi^{j K} + 2 {\mathcal{F}}_{IJ}\bar\vf^I \slashed\partial \varphi^{J} + \frac12  \cF_{IJKL}{}^{kijl}  L_{ij}^I \bar\vf^J_k \slashed{E}^L \varphi_l^K  +  \cF_{IJK}{}{}^{ij} \bar\vf^I_i \slashed{E}^J \varphi^{K}_j    \nn\\
		&&   + \frac12 \cF_{IJK}{}{}^{ij} \bar\vf^J_i \slashed{E}^I \varphi^{K}_j  - 2\cF_{IJKjk} \bar\vf^I_i  \slashed\partial L^{ijJ} \varphi^{kK}   +  \cF_{IJKLk}{}^{ipq} L_{ij}^I   \bar\varphi_p^K \slashed\partial L^{ jkJ} \varphi_q^L \nn\\
		&&  - \frac13 \cF_{IJKL}{}^{ijkl} \bar\vf^I_i  \g^a \vf_j^J \bar\vf_k^K \g_a \vf_l^L    - \frac{1}{12} \cF_{IJKLM}{}^{ijklmn}L_{ij}^I  \bar\vf_k^J \g_a \vf_l^K \bar\vf_m^L \g^a \vf_n^M \,,
		\label{RigidNLinear}
		\eea
		up to partial integrations. Once again, we remind the reader that the index $I$ is fixed and not being summed over. However, if we choose to sum over the $I$ index as well, then the linear multiplet action can be written as
		\bea
		\cL_L &=& -  \frac12 F_{IJ}  \partial_a L_{ij}^I \partial^a L^{ijJ} + \frac14 F_{IJ} E_a^I E^{aJ} - \frac12 F_{IJKj}{}^k E^{abI} \partial_a L^{jl J} \partial_b L^K_{kl}  \,,
		\eea 
		up to an overall minus sign and partial integrations. Here we only provide the bosonic part and made use of the following definitions \cite{deWit4D}
		\bea
		F_{IJ} &=& 2 \cF_{(IJ)} + \cF_{KIJ}{}^{ij} L_{ij}^K \,,\nn\\
		F_{IJK}{}^{ij} &=& 3 \cF_{(IJK)}{}^{ij} + \cF_{LIJK}{}^{klij} L_{kl}^L \,,
		\label{DefFIJ}
		\eea 
		and heavily used the following $SU(2)$ identities \cite{deWit4D}
		\bea
		K_{ik} L^{jk} + K^{jk} L_{ik} &=& \delta_i^j K_{mn} L^{mn} \,,\nn\\
		K_{ij} L_{kl} - K_{kl} L_{ij} &=& \ve_{ik} \ve^{mn} \left(K_{lm} L_{nj} + K_{jm} K_{nl}\right)|_{(i,j) (k,l)} \,.
		\eea
		At this stage, a brief discussion on various choice of $\cF_{IJ}$ is in order
		\begin{enumerate}[i.]
			\item {\underline{$\cF_{IJ} = \delta_{IJ}$:} For this choice, $\cF_{IJ}$ is independent of $L_{ij}$ and all descendants of $\cF_{IJ}$ vanish. The map between the vector and the linear multiplets significantly simplifies, i.e. for a single linear multiplet we have
				\bea
				\O^i =  \slashed\partial \vf^i \,, \qquad Y^{ij} = - \Box L^{ij}   \,, \qquad F_{ab} = - 2 \partial_{[a} E_{b]} \,, 
				\label{SingleLinear1}
				\eea
				and the linear multiplet action is given by
				\bea
				\cL_L &=& -  \partial_a L_{ij} \partial^a L^{ij} + \frac12 E_a E^a - 2 \bar\vf \slashed\partial \vf \,.
				\label{LLinear1}
				\eea
				up to an overall minus sign and boundary terms. The map (\ref{SingleLinear1}) and the action (\ref{LLinear1}) will be particularly useful in the construction of higher derivative linear and vector multiplet actions.
			}
			\item{{\underline{$\cF_{IJ} = \delta_{IJ} (L^I)^{-1}$:} } This choice is particularly relevant to the construction of superconformal linear multiplet actions. In this case, we have
				\bea
				\cF_{IJK}{}^{ij} = - (L^I)^{-3} L^{ij I}\,, \qquad \cF_{IJKL}{}^{ijkl} = (L^I)^{-5} \left(3 L^{ijI} L^{jkI} + \left(L^I \right)^2 
				\ve^{i(k}  \ve^{l)j} \right) \,.
				\eea
				For a single linear multiplet, the bosonic part of Lagrangian that corresponds to this choice is
				\bea
				\cL_{CL} &=& - \frac12 L^{-1} \partial_a L_{ij} \partial^a L^{ij} + \frac14 L^{-1} E_a E^a + \frac12 L^{-3} E^{ab} L^l{}_k \partial_a L^{kp} \partial_b L_{lp} \,,
				\label{ConformalRigid1}
				\eea 
				up to an overall minus sign and boundary terms.
			}
		\end{enumerate}
		
		We may also use the composite expressions for the components of the linear multiplet (\ref{RigidVectorMap}) to construct a vector multiplet action
		\bea
		\cL_V &=& - \frac14 F_{ab} F^{ab} - 2 \bar\O \slashed\partial \O + Y^{ij} Y_{ij}  \,.
		\label{RigidVectorAction}
		\eea
		As with the linear multiplet action, the composite formulae  (\ref{RigidVectorMap}) as well as the vector multiplet action (\ref{RigidVectorAction}) are useful in the construction of higher derivative linear and vector multiplet actions.
		
		
		When coupled to supergravity, $N= (1,0)$ higher derivative vector and linear multiplet actions are known to give rise to higher curvature superinvariants \cite{OzkanThesis,6DHD}. In the case of rigid supersymmetry, we may obtain a higher derivative linear multiplet action by employing the composite vector multiplet fields (\ref{RigidVectorMap}) in the vector multiplet action (\ref{RigidVectorAction})
		\bea
		\cL_L &=& \frac{1}{M^2} \left( \Box L_{ij} \Box L^{ij} - \partial_{[a} E_{b]}\partial^{[a} E^{b]} + 2 \Box \bar\vf \slashed\partial \vf \right) \,,
		\eea
		where $M$ is some mass parameter. We may follow a similar procedure for the vector multiplets and use the composite linear multiplet (\ref{SingleLinear1}) into the linear multiplet action (\ref{LLinear1})
		\bea
		\cL_V &=&  \frac{1}{M^2} \left( -  \partial_a Y_{ij} \partial^a Y^{ij} + \frac12 \partial_b F^{ab} \partial^c F_{ac} - 2 \Box \bar\vf \slashed\partial \vf \right) \,.
		\eea
		One may alternatively wish to use the linear multiplet action (\ref{ConformalRigid1}), however, this Lagrangian involved inverse powers of $L$. To construct a higher derivative vector multiplet action in such a fashion, we first need to consider two distinct linear multiplets represented by $(L_{ij}, \vf_i, E_a)$ and $(L_{ij}^\prime, \vf_i^\prime, E_a^\prime)$ and choose $\cF_{IJ}$ as
		\bea
		\cF_{22} = L^{-1}\,, \qquad \cF_{21}  = - L^{-3} L^\prime_{ij} L^{ij} \,, \qquad \cF_{11} = \cF_{12} = 0 \,.
		\label{NonSymmetricFIJ}
		\eea
		As noted in \cite{Ozkan5D}, such a choice is not symmetric in \textit{(1,2)} indices and satisfies all properties associated with $\cF_{IJ}$. In this case, we may use the primed multiplet as a composite vector multiplet to obtain a higher derivative vector multiplet action. However, such an action also includes the fields of the unprimed linear multiplets.
		
		
		\subsection{Reduction to Five Dimensions}{\label{SSec23}}
		
		The rigid supersymmetric actions for five dimensional $\cN = 2$ vector and linear multiplet actions has already been established for single and multiple number of multiplets \cite{Bergshoeff5D1, Bergshoeff5D2,Ozkan1}. As these multiplets and models can be obtained from a six dimensional $\cN = (1,0)$ theory by a circle reduction, a brief discussion on this issue would be in order before we end our discussion on the rigid six dimensional vector and linear multiplets. The components of the vector multiplet decomposes according to
		\bea
		\left(W_a, \O^i, Y_{ij} \right) \quad  \rightarrow \quad \left( W_{\hat a}, W_5, \l^i, \mathcal{Y}_{ij} \right) \,,
		\eea
		and the linear multiplet decomposes according to
		\bea
		\left( L_{ij}, \vf^i, E_a \right) \quad  \rightarrow \quad \left( {\rm{L}_{ij}}, \hat\vf^i, {\rm{E}}_{\hat a} , {\rm{E}}_5 \right) \,.
		\eea
		where $E^{\hat a}$ is the four-dimensional divergence-free vector field, i.e. $\partial^{\hat a} {\rm{E}}_{\hat a} = 0$. To be more precise, let us introduce our convention for gamma-matrices, supersymmetry parameter $\e$ and Dirac conjugated spinors \cite{BergshoeffSezgin5D}
		\bea
		\gamma_a = \rmi \Gamma_{\hat a} \Gamma_5 \,, \qquad \e = \ve\,, \qquad \bar\epsilon = \rmi \bar\ve \Gamma_5 \,.
		\eea
		We now make the following ansatz for the fields of the five dimensional $\cN = 2$ vector multiplet \cite{BergshoeffSezgin5D}
		\bea
		W_a = A_{\hat a}\,, \qquad W_5 = \rho \,, \qquad Y_{ij} = - {\mathcal Y}_{ij}  \,, \qquad \O =  -\frac12 \l 
		\eea
		and obtain the supersymmetry transformation rules as
		\begin{align}
			\d  \r =& \frac12 \rmi \bar\ve \l \,, &  \d A_{\hat \m} =& \frac12 \bar\ve \G_{\hat\m} \l  \,,\nn\\
			\d \l^i =& - \frac14 \G \cdot F \ve^i - \frac12 \rmi \slashed\partial \rho \ve^i - {\mathcal Y}^{ij} \ve_j\,, & \d {\mathcal Y}^{ij} =& -\frac12 \ve^{(i} \slashed\partial \l^{j)}
		\end{align}
		These transformation rules precisely match with the ones given in \cite{Bergshoeff5D1}. For the fields of the linear multiplet, we make the following ansatz
		\bea
		L_{ij} = {\rm L}_{ij} \,, \qquad \vf^i = \G_5 \hat\vf^i \,, \qquad E_a = -2 \rm{E}_{\hat a}\,, \qquad E_5 = - 2 N \,,
		\eea
		which gives rise to the following supersymmetry transformation rules
		\begin{align}
			\d \rm{L}^{ij} =& \rmi \bar\ve^{(i} \hat\vf^{j)} \,, & \d \hat{\vf}^i =& - \frac12 \rmi \slashed\partial \rm{L}^{ij} \ve_j - \frac 12 \rmi \slashed{\rm{E}} \ve^i  \,,\nn\\
			\d {\rm{E}}_{\hat a} =& -\frac12 \rmi \bar\ve \Gamma_{\hat a \hat b} \partial^{\hat b} \hat\vf \,, & \d N =& \frac12 \bar\ve  \slashed\partial \hat\vf \,.
		\end{align}
		These transformation rules precisely match with \cite{Bergshoeff5D1} and its application to supersymmetric linear multiplet Lagrangian (\ref{RigidNLinear}) reproduces the five dimensional  linear multiplet actions \cite{Ozkan5D}.
		

		\subsection{Superconformal Vector and Linear Multiplets}{\label{SSec24}}
		
		So far we only considered vector and linear multiplets that are representation of six-dimensional $\cN= (1,0)$ super-Poincar\'e algebra. However, these multiplets can be assigned full $\cN=(1,0)$ superconformal symmetry in which case the fields pick up additional dilatation $(\L_D)$ and special supersymmetry $(\eta)$ transformations. For the linear multiplet, the transformation rules are given by
		\bea
		\delta L^{ij} &=&  \bar{\epsilon}^{(i} \varphi^{j)} + 4 \L_D L^{ij} \nonumber \,,\\
		\delta \varphi^i &=& \frac12 \slashed{\partial} L^{ij} \epsilon_j - \frac14 \slashed{E} \epsilon^i - 4 L^{ij} \eta_j  + \frac92 \L_D \vf^i  \nonumber\,,\\
		\delta E_{a} &=& \bar\epsilon \gamma_{ab}\partial^b \varphi - 5 \bar\eta \g_a \vf + 5 \L_D E_a \,,
		\label{LinearRigidConformalSUSY}
		\eea
		and for the vector multiplet, the transformation rules are
		\bea
		\delta W_\mu &=& - \bar{\epsilon} \gamma_\mu \Omega \,,\nn\\
		\delta \Omega^i &=& \frac{1}{8} \gamma \cdot F \epsilon^i - \frac12 Y^{ij} \epsilon_j + \frac32 \L_D \Omega^i  \,,\nn\\
		\delta Y^{ij} &=& - \bar{\epsilon}^{(i} \slashed\partial \Omega^{j)}  + 2 \bar\eta^{(i} \Omega^{j)}  + 2 \L_D Y^{ij} \,.
		\label{VectorRigidConformalSUSY}
		\eea
		These additional symmetries have immediate implications on the function $\cF_{IJ}$. First of all, the scaling dimension of $\cF_{IJ}$ must be $-4$ in order to match the scaling dimension of both sides in the composite formulae (\ref{CompositeOmega}) and (\ref{RigidYandF}) 
		\bea
		\d_D \cF_{IJ}(L) = - 4 \L_D \cF_{IJ}(L)\,.
		\label{ScalingF}
		\eea
		As a result, the choice $\cF_{IJ} = \d_{IJ}$ no longer holds in the superconformal case, $\cF_{IJ} = \d_{IJ} (L^I)^{-1}$ or the non-symmetric choice (\ref{NonSymmetricFIJ}) are still valid. Next, we turn to the special supersymmetry transformations of the fields. As $\O^i$ is S-SUSY invariant, $\cF_{IJ}$ must satisfy
		\bea
		\cF_{IJKik}L^{kjK} = -  \frac12 \,\d_i^j \, \cF_{IJ} \,.
		\eea
		We may multiply this identity with $L^{km J}$ and obtain a useful result \cite{deWit4D}
		\bea
		\cF_{IJKij} L_{pq}^{J} L^{pqK} = - \cF_{IJ} L^J_{ij} \,,
		\eea
		which further implies that
		\bea
		\cF_{IJ} L_{ij}^I L^{ijJ} = F_{IJ} L_{ij}^I L^{ijJ}  \,.
		\eea
		To summarize, the conformal $\cF_{IJ}$ needs to satisfy the following four identities
		\begin{align}
			\cF_{IJK}{}^{ij} &= \cF_{I(JK)}{}^{ij} \,, & \cF_{IJKi[jk]l} &= 0 \,,\nn\\
			\d_D \cF_{IJ} = &  - 4 \L_D \cF_{IJ}(L)\,, & 	\cF_{IJKik}L^{kjK} &= -  \frac12 \,\d_i^j \, \cF_{IJ} \,.
			\label{ConformalF}
		\end{align}
		
		While the composite superconformal vector multiplet can be achieved by the proper choice of $\cF_{IJ}$, the composite linear multiplet (\ref{RigidVectorMap}) immediately fails due to the scaling dimension of the fields. However, one may push the idea that a superconformal composite linear multiplet can be obtained by mapping a vector as well as a compensating linear multiplet. For instance, one can start with the following ansatz for the $SU(2)$ scalar of a primed vector multiplet $L_{ij}^\prime$
		\bea
		L_{ij I}^\prime &=& {\mathcal{M}}_{IJ} Y_{ij}^J + {\mathcal M}_{IJKk(i} \bar\vf^{kK} \O_{j)}^J \,.
		\eea
		where $\cM_{IJ} $ is a function of $L_{ij}$ i.e. $\cM_{IJ}= \cM_{IJ}(L)$. By demanding proper dilatation, S- and Q-SUSY transformations, it is possible to push this ansatz to all fields and find the constraints that the function ${\mathcal{M}}_{IJ}$ should satisfy. However, as we will see in the next section, there is a simpler possibility to achieve a superconformal composite linear multiplet when we couple to conformal supergravity. The  ``dilaton Weyl multiplet" of $\cN=(1,0)$ superconformal theory includes a dilaton field $\sigma$ and a dilatino field $\p^i$ that we can use to compensate the dilatation and S-SUSY transformations of the superconformal completion of the map (\ref{RigidVectorMap}).

		\section{Conformal Supergravity}{\label{Sec3}}
		
		In the previous section, we discussed the superconformal transformations of linear and vector multiplets that are defined in flat space where we dealt with space-time independent transformation parameters. In a superconformal background, where the rigid parameters are replaced by space-time ones, the transformation rules contain the gauge fields of the superconformal theory. In general, however, the gauge fields do not have the right counting to form a closed background ``Weyl" multiplet but the inclusion of matter fields is necessary. For the six dimensional $\cN = (1,0)$ theory there are two sets of possible matter fields, one leading to the so-called standard Weyl multiplet and the other leading to the dilaton Weyl multiplet. We defer the details of these multiplets to the Appendix \ref{Appendix1}. The $Q-$ and $S-$supersymmetry transformation rules for the linear multiplet are given by \cite{Bergshoeff6D,Coomans6D}
		\bea
		\delta L^{ij} &=&  \bar{\epsilon}^{(i} \varphi^{j)}  \nonumber \,,\\
		\delta \varphi^i &=& \frac12 \slashed{\cD} L^{ij} \epsilon_j - \frac14 \slashed{E} \epsilon^i - 4 L^{ij} \eta_j   \nonumber\,,\\
		\delta E_{a} &=& \bar\epsilon \gamma_{ab}\cD^b \varphi + \frac{1}{24} \bar\e \g_a  \cdot T^- \vf - \frac13 \bar\e^i \g_a \chi^j L_{ij} - 5 \bar\eta \g_a \vf \,,
		\label{LinearLocalConformalSUSY}
		\eea
		where the superconformal covariant derivatives are defined as
		\bea
		\cD_\m L^{ij} &=& \partial_\m L^{ij} - 4 b_\m L^{ij}  + 2 \cV_\m{}^{(i}{}_k L^{j)k} - \bar\p_\m^{(i} \, \vf^{j)}\,,\nn\\
		\cD_{\m} \vf^i &=& \partial_\m \vf^i - \frac92 b_\m \vf^i - \cV_\m^{ij} \vf_j - \frac12 \slashed\cD L^{ij} \p_{\m j} + \frac14 \slashed{E} \p_\m^i + 4 L^{ij} \phi_j \,.
		\eea
		As in the rigid case, the algebra closes if $E_a$ satisfies \cite{Bergshoeff6D,Coomans6D}
		\bea
		\cD^a E_a - \frac12 \bar\vf \chi =0 \,,
		\label{ConstraintE}
		\eea
		where the superconformal covariant derivative of $E_a$ is defined as
		\bea
		\cD_\m E_a &=& \partial_\m E_a - 5 b_\m E_a + \o_{\m a}{}^b E_b - \bar\p_\m \g_{ab} \cD^b \vf-  \frac{1}{24} \bar\p_\m \g_a  \g \cdot T^- \vf \nn\\
		&&+ \frac13 \bar\p^i_\m \g_a \chi^j L_{ij}  + 5 \bar\phi_\m \g_a \vf \,.
		\eea
		Note that the constraint equation (\ref{ConstraintE}) requires an additional $\bar\vf \chi$ term in order to maintain the $Q-$ and $S-$invariance of the constraint. The constraint on $E_a$ allow us to define a four-form gauge field $E_{\m\n\r\s}$ which can be dualized to a two-form gauge field $E_{\m\n}$ that is defined via
		\bea
		E^a = e_\m{}^a \cD_\n E^{\m\n} \,,
		\eea
		where the supersymmetry transformation rule for $E_{\m\n}$ is given by
		\bea
		\d E_{\m\n} &=& \bar\e \g_{\m\n} \vf + \bar\vf^{\r}_i \g_{\m\n\r} \e_j L^{ij} \,.
		\eea
		The gauge fields of the Weyl multiplets that appear here are the sechsbein $e_\m{}^a$, the spin-connection $\o_\m{}^{ab}$, the dilatation gauge field $b_\mu$, the $SU(2)$ R-symmetry gauge field $\cV_\m^{ij}$, and the $Q-$ and the $S-$ supersymmetry gauge fields  $\p_\m^i$ and $\phi_\m^i$. As the matter multiplets are inert under special conformal symmetry, its corresponding gauge field $f_\m{}^a$ does not appear in the transformation rules. The set of fields $(\o_\m{}^{ab},f_\m{}^a, \phi_\m^i)$ are not independent but can be expressed in terms of the independent fields $(e_\m{}^a, b_\m, \cV_\m^{ij}, \p_\m^i)$. Transformation rules also consist of a real scalar field $D$, an antisymmetric tensor of negative duality $T_{abc}^-$ and an $SU(2)$ Majorana-Weyl spinor of negative chirality $\chi^i$. These matter fields are the fundamental fields of the standard Weyl multiplet but can be expressed in terms of a dilaton Weyl multiplet coupled to a tensor multiplet. For the vector multiplet, the $Q-$ and $S-$ transformation rules are given by  \cite{Bergshoeff6D,Coomans6D}
		\bea
		\delta W_\mu &=& - \bar{\epsilon} \gamma_\mu \Omega \,,\nn\\
		\delta \Omega^i &=& \frac{1}{8} \gamma \cdot \widehat F \epsilon^i - \frac12 Y^{ij} \epsilon_j  \,,\nn\\
		\delta Y^{ij} &=& - \bar{\epsilon}^{(i} \slashed\cD \Omega^{j)}  + 2 \bar\eta^{(i} \Omega^{j)}   \,,
		\label{VectorLocalConformalSUSY}
		\eea
		where the superconformal field strength $\widehat F_{\m\n}(W)$ and the supercovariant derivative $\cD_\m \O^i$ are defined as
		\bea
		\widehat F_{\m\n}(W) &=& 2 \partial_{[\m} W_{\n]} + 2 \bar\p_{[\m} \g_{\n]} \O \,,\nn\\
		\cD_\m \O^i &=& \partial_\m \O^i - \frac32 b_\m \O^i + \frac14 \o_\m{}^{ab} \g_{ab} \O^i - \frac12 \cV_{\m}{}^i{}_j \O^j - \frac18 \g \cdot \widehat F \p_\m^i + \frac12 Y^{ij} \p_{\m j} \,.
		\eea	
		With the supersymmetry transformation rules in hand, we are now in a position to generalize rigid composite vector multiplet (\ref{CompositeOmega}) and (\ref{RigidYandF}) and to construct local linear and vector multiplet actions. The starting point of an action principle is the generalization of rigid auxiliary action (\ref{VectorLinearAction}) with the inclusion of the Weyl multiplet fields \cite{Bergshoeff6D,Coomans6D}
		\bea
		e^{-1} \cL_{VL} = Y^{ij} L_{ij} + 2 \bar\O \vf - L_{ij} \bar\p_{\m}^i \g^\m \O_j + \frac14 F^{\m\n} E_{\m\n} \,.
		\label{LocalVL}
		\eea
		Given the properties (\ref{ScalingF}) and (\ref{ConformalF}), the composite vector multiplet is given by
		\bea
		\Omega^i_I &=& {\mathcal{F}}_{IJ}\slashed \cD \varphi^{iJ} + \cF_{IJKjk} \slashed\cD L^{ijJ} \varphi^{kK} + \frac23 \cF_{IJ} L^{ij J} \chi_j  + \frac{1}{12} \cF_{ij} \g \cdot T^- \vf^{iJ} - \frac12 \cF_{IJK}{}{}^{ij} \slashed{E}^J \varphi^{K}_j \nn\\
		&&  + \frac16 \cF_{IJKL}{}^{ijkl} \g^a \vf_j^J \bar\vf_k^K \g_a \vf_l^L  \,,\nn\\
		Y^{ij}_{I}&=& - \cF_{IJ} \Box^c L^{ijJ} - \cF_{IJKkl} \cD_a L^{ik J} \cD^a L^{jl K} - \cF_{IJKk}{}^{(i} \cD_a L^{j)k J} E^{a K} - \frac14 \cF_{IJK}{}^{ij} E_a^J E^{a K} \,,\nn\\
		&& - \frac13 \cF_{IJ} L^{ij J} D + \frac16 \cF_{IJ} \bar\chi^{(i} \vf^{j)J} + \frac43 \cF_{IJK}{}^{k(i} L^{j)lJ} \bar\chi_k \vf_l^K  + \frac{1}{12} \cF_{IJK}{}^{ij} \bar\vf \g \cdot T^- \vf \nn\\ 
		&& - \frac14 \cF_{IJK}{}^{ij} \bar\varphi^J \slashed\cD \varphi^K - 2 \cF_{IJK}{}^{k(i} \bar\vf_k^J \slashed \cD \varphi^{j) K} -  \cF_{IJKL}{}^{pqk(i} \cD_a L^{J j)}{}_k \bar\varphi_p^K \gamma^a \varphi_q^L \nn\\
		&& + \frac12  \cF_{IJKL}{}^{k(ij)l}  \bar\vf^J_k \g^a \varphi_l^K  E_a^L - \frac{1}{12} \cF_{IJKLM}{}^{ijklmn}\bar\vf_k^J \g_a \vf_l^K \bar\vf_m^L \g^a \vf_n^M \,,\nn\\ 
		\widehat F_{ab I}&=&  - 2 \cD_{[a} \left(  \cF_{IJ} E_{b]}^J  \right) + 2 \cF_{IJ} L^{ijJ} \widehat{R}_{abij}(\cV)+ 2 \cF_{IJKj}{}^k \cD_{[a} L^{jl J} \cD_{b]} L_{lk}^K  \nn\\
		&& + \cF_{IJ} \bar{\widehat{R}}_{ab}(Q) \vf^J + 2 \cD_{[a} \left(\cF_{IJK}{}^{ij} \bar\vf_i^J \g_{b]} \varphi_j^K\right)\,.
		\label{ConformalCompositeVector}
		\eea
		where $\widehat R_{\m\n}{}^{ij} (\cV)$ and $\widehat R_{\m\n}{}^i (Q)$ are the superconformal covariant curvatures of the gauge fields $\cV_{\m}{}^{ij}$ and $\p_\m^i$ respectively, see Appendix \ref{Appendix1}. The superconformal d'Alembertian $\Box^c L_{ij}^I$ is defined as
		\bea
		\Box^c L^{ij}_I &=& \left(\partial^a - 5 b^a + \o_b{}^{ba} \right) \cD_a L^{ij}_I - 8 f_a{}^a L^{ij}_I + 2 V^{\m(i}{}_k \cD_\m L^{j)k}_I - \bar\p_\m^{(i} \cD^\m \vf^{j)}_I- \frac16 \bar\p_\m \g^\m \chi L^{ij}_I  \nn\\
		&& + \frac{1}{6} \bar\p_\m^{(i} \g^\m \chi^{k)} L^{j}{}_{kI} + \frac{1}{6} \bar\p_\m^{(j} \g^\m \chi^{k)} L^{i}{}_{kI} - \frac1{24} \bar\vf^{(i}_I \g \cdot T^- \g^\m \p_{\m}^{j)} - \bar\vf_I^{(i} \g^\m \phi_\m^{j)} \,.
		\eea 
		As mentioned in the previous section, it is not possible to construct a conformal action for the vector multiplet unless we use a compensating multiplet. The compensating multiplet can be chosen as a linear multiplet, however, a local vector multiplet action then contains the fields of vector, linear as well as a Weyl multiplet. From a more minimalist approach, it is possible to use the dilaton Weyl multiplet to compensate the conformal symmetries. In that case, the correct transformation rule for a composite $L_{ij}$ can be obtained by using the scalar field of the dilaton Weyl multiplet $\s$ as well as the dilatino field $\p^i$ as extra fields \cite{Bergshoeff6D}
		\bea
		L^{ij} &=& \s Y^{ij} + 2 \bar\p^{(i} \O^{j)} \,.
		\label{CompositeLocalLinear1}
		\eea
		The rest of the composite linear multiplet can then be obtained by the $Q-$variation of the composite $L^{ij}$, which is given by \cite{Bergshoeff6D}
		\bea
		\vf^i &=& - \s \slashed\cD \O^i - \frac12 \slashed\cD \s\, \O^i + \frac{1}{24} \g \cdot H\, \O^i + \frac14 \g \cdot \widehat F\, \p^i + Y^{ij} \p_j \,,\nn\\
		E_{\m\n} &=& - \s F_{\m\n} - \frac14  \e_{\m\n\r\s\l\t} B^{\r\s}  F^{\l\t}  + 2 \bar\O \g_{\m\n} \p + \s \bar\O \g^\r \g_{\m\n} \p_\r   \,,
		\label{CompositeLocalLinear2}
		\eea
		where $H_{\m\n\r}$ is the field strength of two-form gauge field $B_{\m\n}$, which is one of the matter fields in the dilaton Weyl multiplet
		\bea
		H_{\m\n\r} = 3 \partial_{[\m} B_{\n\r]}  +  3 \bar\p_{[\m} \g_{\n\r]} \p + \frac32 \s \bar\p_{[\m} \g_\n \p_{\r]} \,.
		\eea
		With the composite multiplets and the action principle (\ref{LocalVL}) it is possible to construct local superconformal actions for vector and linear multiplets. For the linear multiplet, the action contains a Ricci scalar term via the composite $Y_{ij}$ due to superconformal d'Alembertial of $L_{ij}$. This action can therefore be used to express an off-shell Poincar\'e supergravity after gauge fixing. Noticing that the bosonic part of composite $\widehat F_{ab I}$ can be written as
		\bea
		F_{ab I} &=& -  2 \partial_{[a} \left( \cF_{IJ} E_{b]}^J - 2 \cF_{IJ} L_{ij}^J \cV_{b]}{}^{ij} \right) +  2 \cF_{IJKj}{}^k \partial_{[a} L^{jl J} \partial_{b]} L_{lk}^K \,,
		\eea
		we give the bosonic part of the linear multiplet action for $n$-number of linear multiplets as
		\bea
		e^{-1} \cL_L &=& - \cF_{IJ} L_{ij}^I \Box^c L^{ijJ} - \cF_{IJKkl} L_{ij}^I  D_a L^{ik J} D^a L^{jl K} - \cF_{IJKk}{}^{i} L_{ij}^I  D_a L^{jk J} E^{a K}  \,,\nn\\
		&& - \frac14 \cF_{IJK}{}^{ij} L_{ij}^I E_a^J E^{a K} - \frac13 \cF_{IJ} L_{ij}^I  L^{ij J} D  -  \frac12  \cF_{IJ} E^{aI} E_{a}^J  + \cF_{IJ} E^{a I} L_{ij}^J \cV_{a}^{ij} \nn\\
		&& + \frac12 \cF_{IJKj}{}^k E^{ab I}  \partial_{a} L^{jl J} \partial_{b} L_{lk}^K \,,
		\label{LocalLinear}
		\eea
		where the $SU(2)$ covariant derivative is defined as
		\bea
		D_\m L^{ij} &=& \partial_\m L^{ij}  + 2 \cV_\m{}^{(i}{}_k L^{j)k} \,.
		\eea
		As before, there is no sum in the $I$ index and no particular symmetry in $(I,J)$ indices. However, if we choose to sum over the index $I$, then the bosonic part of the supersymmetric action considerably simplifies
		\bea
		e^{-1} \cL_{L, SW} &=& \frac25 F_{IJ} L_{ij}^I L^{ij J} R   -  \frac2{15} F_{IJ} L_{ij}^I L^{ij J} D -  \frac12 F_{IJ}  D_a L_{ij}^I D^a L^{ijJ} + \frac14 F_{IJ} E_a^I E^{aJ}  \,,\nn\\
		&&  + F_{IJ} E^{a I} L_{ij}^J \cV_{a}^{ij} - \frac12 F_{IJKj}{}^k E^{abI} \partial_a L^{jl J} \partial_b L^K_{kl}   \,,
		\eea 
		where $F_{IJ}$ and its descendants are as defined (\ref{DefFIJ}) and $R$ is the Ricci scalar. There are three notable features of this action. First, as the linear multiplet is inert under special conformal transformations, $b_\m$, the gauge field of dilatations, is the only independent field that transforms non-trivially under special conformal symmetry. As a result, all $b_\mu$ terms cancel each other out and the action does not contain a $b_\m$ term. Second, the scalar field $D$ imposes a severe constraint on the scalars of the linear multiplet, i.e. $F_{IJ} L_{ij}^I L^{ij J}$. Such a constraint also annihilates the pre-factor of the Ricci scalar, not allowing us to obtain off-shell Einstein-Hilbert supergravity after gauge fixing. Third, as can be seen from the composite formulate (\ref{ConformalCompositeVector}), the field equation for $T_{abc}^-$ imposes a constraint on the fermionic fields. Thus, we use the map between the standard and the dilaton Weyl multiplets, see Appendix \ref{Appendix1}, and establish the linear multiplet action in a dilaton Weyl multiplet background
		\bea
		e^{-1} \cL_{L, DW} &=& \frac12 F_{IJ} L_{ij}^I L^{ij J} R   - \frac12  \s^{-1}  F_{IJ} L_{ij}^I L^{ij J}\Box \s - \frac1{24} \s^{-2} F_{IJ} L_{ij}^I L^{ij J}  H_{abc} H^{abc} \nn\\
		&& -  \frac12 F_{IJ}  D_a L_{ij}^I D^a L^{ijJ} + \frac14 F_{IJ} E_a^I E^{aJ}   + F_{IJ} E^{a I} L_{ij}^J \cV_{a}^{ij} \nn\\
		&& - \frac12 F_{IJKj}{}^k E^{abI} \partial_a L^{jl J} \partial_b L^K_{kl}  \,.
		\label{ConformalLinDW}
		\eea
		As we will discuss in detail, the presence of two scalar fields, $\s$ and $L_{ij}$, allow us two distinct gauge fixing possibilities, one giving rise to an off-shell supergravity in the Jordan frame and the other in Einstein frame. Deferring this discussion to the next section, we give the bosonic part of the conformal vector multiplet action
		\bea
		e^{-1} \cL_{V} &=& -\frac14 \s F^{ab} F_{ab} - \frac1{16} \e^{abcdef} B_{ab} F_{cd} F_{ef} +\s Y^{ij} Y_{ij} \,,
		\eea
		where we use the composite linear multiplet (\ref{CompositeLocalLinear1}) and (\ref{CompositeLocalLinear2}) as well as the local vector-linear action (\ref{LocalVL}). Finally, for future reference, we take the vector multiplet in the local vector-linear action and the composite linear multiplet to be different, which gives rise to a vector multiplet action for two vector multiplets
		\bea
		e^{-1} \cL_{V V^\prime} &=& -\frac14 \s F^{ab} F_{ab}^\prime - \frac1{16} \e^{abcdef} B_{ab} F_{cd} F_{ef}^\prime +\s Y^{ij} Y_{ij}^\prime \,,
		\label{VVprime}
		\eea
		where the second multiplet is expressed by the primed quantities $(\Omega_i^\prime, W_\mu^\prime, Y_{ij}^\prime)$.

		\section{Off-Shell Supergravity Models with Multiple Linear Multiplets}{\label{Sec4}}
		
		In this section, we take advantage of the conformal linear and vector multiplet actions to construct various off-shell supergravity models. First, we use the superconformal linear multiplet action in the dilaton Weyl multiplet background (\ref{ConformalLinDW}) to obtain an off-shell description of six dimensional $\cN = (1,0)$ supergravity. This can be achieved in two ways, one giving rise to Einstein-frame model and the other giving rise to a Jordan-frame model, depending on the gauge fixing condition. Next, we eliminate the auxiliary fields and present the on-shell supergravity coupled to $n$-number of linear multiplets. Finally, we discuss higher derivative models where the leading terms are not higher-order curvature terms but curvature terms coupled to auxiliary fields. When coupled to Poincar\'e supergravity, off-diagonal invariants give rise to on-shell higher curvature supergravity models after the elimination of auxiliary fields.
		
		
		\subsection{Poincar\'e Supergravity}{\label{Sec41}}
		
		An off-shell Poincar\'e supergravity can be obtained from the conformal linear multiplet action (\ref{ConformalLinDW}) by gauge fixing the redundant dilatation, special conformal and special supersymmetry transformation,  see Table \ref{Table}. It is also possible to break the $SU(2)$ R-symmetry to a $U(1)$ subgroup.
		
		\subsubsection*{Einstein-frame Off-Shell Supergravity}
		
		If we consider a single linear multiplet coupled to dilaton Weyl multiplet, which corresponds to setting $\cF_{11} = L^{-1}$ in (\ref{ConformalLinDW}), we may adopt the following gauge fixing conditions
		\bea
		b_\mu = 0 \,, \qquad L_{ij} = \frac1{\sqrt 2} \delta_{ij} L\,, \qquad L = 1\,, \qquad \varphi^i = 0\,,
		\label{GaugeFixing1}
		\eea
		\begin{table}[h!]
			\centering
			\begin{tabular}{|c|c|c|c|}
				\hline
				\textbf{Weyl Multiplet}                                                                                                                                         & \textbf{Compensator}                                                                           & \textbf{Gauge Fixing}                                                                                                                & \textbf{Poincar\'e Multiplet        }                                                                                                                                                      \\ \hline
				\begin{tabular}[c]{@{}c@{}}Dilaton Weyl Multiplet\\ $(e_\mu{}^a, \psi_\mu^i, b_\mu, {\mathcal V}_\mu{}^{ij}, \sigma, B_{\mu\nu}, \psi^i)$\end{tabular} & \begin{tabular}[c]{@{}c@{}}Linear Multiplet\\ $(L_{ij}, \varphi^i, E_a)$\end{tabular} & \begin{tabular}[c]{@{}c@{}}$b_\mu = 0$\\ $L_{ij} = \frac1{\sqrt 2} \delta_{ij} L$\\ $L= 1$\\ $\varphi^i = 0$\end{tabular}   & \begin{tabular}[c]{@{}c@{}}$(e_\mu{}^a, \psi_\mu^i, b_\mu, {\mathcal V}_\mu,{\mathcal V}^\prime_\mu{}^{ij}$\\ $\sigma, B_{\mu\nu}, \psi^i, E_a)$\\ \\ (Einstein-frame)\end{tabular} \\ \hline
				\begin{tabular}[c]{@{}c@{}}Dilaton Weyl Multiplet\\ $(e_\mu{}^a, \psi_\mu^i, b_\mu, {\mathcal V}_\mu{}^{ij}, \sigma, B_{\mu\nu}, \psi^i)$\end{tabular} & \begin{tabular}[c]{@{}c@{}}Linear Multiplet\\ $(L_{ij}, \varphi^i, E_a)$\end{tabular} & \begin{tabular}[c]{@{}c@{}}$b_\mu = 0$\\ $L_{ij} = \frac1{\sqrt 2} \delta_{ij} L$\\ $\sigma= 1$\\ $\psi^i = 0$\end{tabular} & \begin{tabular}[c]{@{}c@{}}$(e_\mu{}^a, \psi_\mu^i, b_\mu, {\mathcal V}_\mu, {\mathcal V}^\prime_\mu{}^{ij}$\\ $L, B_{\mu\nu}, \varphi^i, E_a)$\\ \\ (Jordan-frame)\end{tabular}  \\ \hline
			\end{tabular} 
			\caption{List of possible off-shell supergravity construction via gauge fixing. In a standard Weyl multiplet background, it is not possible to obtain a consistent on-shell Einstein-Hilbert supergravity. In a Weyl multiplet background, with a single linear multiplet coupling, a gauge fixing condition that utilizes the scalar field of dilaton Weyl multiplet leads to a Jordan-frame off-shell supergravity while a gauge choice that uses the fields of linear multiplet leads to Einstein-frame supergravity.}
			\label{Table}
		\end{table}
		where the first choice fixes special conformal symmetry, the second one breaks the $SU(2)$ R-symmetry to $U(1)$, the third breaks the dilatation symmetry and the last one fixes the special supersymmetry. As a result, we obtain an off-shell Poincar\'e supergravity in Einstein frame \cite{Bergshoeff6D,Coomans6D}
		\bea
		e^{-1} \cL = \frac{1}{2} R - \frac14 E^a E_a + \cV^\prime_{a ij} \cV^{\prime ij}_a + \frac{1}{\sqrt2} E^a \cV^\prime_a{}^{ij} \d_{ij} - \frac12 \s^{-2} \partial_\m \s \partial^\m \s - \frac{1}{24} \s^{-2} H_{\m\n\r} H^{\m\n\r}  \,,
		\label{EHinEF}
		\eea
		where we only provide the bosonic part. Due to our gauge choice that breaks the $SU(2)$ R-symmetry, we also decompose the $SU(2)$ R-symmetry gauge field $\cV_{a}^{ij}$ into its trace and traceless part
		\bea
		\cV_a = \cV_a{}^{ij} \d_{ij} \,, \qquad \cV^\prime_a{}^{ij} = \cV_a{}^{ij} - \frac12 \delta_{ij} \cV_a{}^{kl} \d_{kl} \,.
		\eea
		When $n$-linear multiplet coupling is considered, we may adopt the following gauge fixing conditions to obtain the supergravity coupling of $(n-1)$-linear multiplets to off-shell supergravity in Einstein frame
		\bea
		b_\mu = 0 \,, \qquad F_{IJ} L_{ij}^I L^{ij J} = 1 \,, \qquad F_{IJK}{}^{ij} \vf_j^K L_{mn}^I L^{mn J} + 2 F_{IJ} L^{ij I} \vf_j^J = 0 \,.
		\label{GaugeFixing2}
		\eea
		where the second choice fixes the dilatation symmetry while the last one fixes the special supersymmetry. As a result, we obtain an off-shell action for linear multiplets
		\bea
		e^{-1} \cL &=& \frac12 R - \frac12 \s^{-2} \partial_\m \s \partial^\m \s - \frac{1}{24} \s^{-2} H_{\m\n\r} H^{\m\n\r}-  \frac12 F_{IJ}  D_a L_{ij}^I D^a L^{ijJ} \nn\\
		&& + \frac14 F_{IJ} E_a^I E^{aJ}   + F_{IJ} E^{a I} L_{ij}^J \cV_{a}^{ij}  - \frac12 F_{IJKj}{}^k E^{abI} \partial_a L^{jl J} \partial_b L^K_{kl} \,,
		\label{EHinJF}
		\eea
		where the scalars $L_{ij}^I$ and the fermions $\vf^I$ are restricted by the equation (\ref{GaugeFixing2}).	
		
		\subsubsection*{Jordan-frame Off-Shell Supergravity} 
		
		We may use the scalar of the dilaton Weyl multiplet $\s$ to fix the dilatation symmetry. In this case, the consistent set of gauge fixing condition is given by \cite{BergshoeffHD}
		\bea
		b_\m = 0 \,, \qquad \s = 1 \,, \qquad \psi^i = 0 \,,
		\label{GaugeFixing3}
		\eea
		where the first choice fixes special conformal symmetry, the second one fixes the dilatations and the third one fixes special supersymmetry. In this case, the off-shell supersymmetric action is given by
		\bea
		e^{-1} \cL &=& g_{{\rm Lin}} \left(\frac12 R - \frac{1}{24} H_{\m\n\r} H^{\m\n\r} \right) -  \frac12 F_{IJ}  D_a L_{ij}^I D^a L^{ijJ} + \frac14 F_{IJ} E_a^I E^{aJ}    \nn\\
		&& + F_{IJ} E^{a I} L_{ij}^J \cV_{a}^{ij} - \frac12 F_{IJKj}{}^k E^{abI} \partial_a L^{jl J} \partial_b L^K_{kl}  \,,
		\label{EHinJF2}
		\eea
		where the potential $g_{Lin} (L) \equiv g_{\rm Lin}$ is defined by
		\bea
		g_{Lin} = F_{IJ} L_{ij}^I L^{ij J} \,.
		\eea


		\subsection{Off-Diagonal $R Y_{ij}$ Invariant and $R^2$ Supergravity}{\label{Sec42}}
		
		When higher curvature supergravity models are needed, the most straightforward strategy is to construct off-shell models, if possible, then add the off-shell higher curvature models to the off-shell Poincar\'e supergravity and perturbatively eliminate the auxiliary fields order by order in the small perturbation parameter, see i.e. \cite{6DHD,Cremonini5D,OzkanThesis} for six and five dimensional examples. This approach has been widely used in various dimensions for a various number of supercharges. The supersymmetric higher curvature models usually include coupling between the auxiliary field of the off-shell Poincar\'e supergravity and the curvature terms. Then, the elimination of auxiliary fields leads to gravitational higher derivative terms, spoiling the particular combination one is after. Off-diagonal invariants are supersymmetric off-shell models where the leading terms are not higher-order curvature terms but curvature terms coupled to auxiliary fields. Previously, these models have been used to eliminate such undesired couplings \cite{SuperBHT,BISugra,3DN2}. Here, we aim to construct an off-diagonal invariant that leads to an on-shell $R^2$-supergravity (with a coupling to vector multiplet) which can be compared with \cite{OzkanThesis,6DHD}.
		
		We start with a composite superconformal vector multiplet that is achieved according to a single conformal truncation, i.e.
		\bea
		\cF_{11} = L^{-1} \,, \qquad \cF_{111}{}^{ij} = - L^{-3} L^{ij} \,.
		\label{SingleMultipletTruncation}
		\eea
		Upon gauge fixing (\ref{GaugeFixing3}), the composite off-shell vector multiplet, $(\Omega_i^\prime, W_\mu^\prime, Y_{ij}^\prime)$, is given by
		\bea
		Y^{\prime ij} &=& - L^{-1} D^a D_a L^{ij} + \frac12 L^{-1} L^{ij}R - \frac1{24} L^{-1} L^{ij} H_{abc}H^{abc} + L^{-3} L_{kl} D_a L^{ik} D^a L^{jl}\nn\\
		&&+ L^{-3} E^a L_k{}^{(i} D_a L^{j)l} + \frac14 L^{-3} L^{ij} E^a E_a \,,\nn\\
		F_{ab}^\prime &=& -  2 \partial_{[a} \left( L^{-1} E_{b]} - 2 L^{-1} L_{ij} \cV_{b]}{}^{ij} \right) -  2 L^{-3} L_{j}{}^k \partial_{[a} L^{jl} \partial_{b]} L_{lk} \,.
		\eea
		where 
		\bea
		D^a D_a L^{ij} &=& (\partial^a + \o_b{}^{ba} ) D_a L^{ij} + 2 \cV^{a(i}{}_k D_a L^{j)k} \,.
		\eea
		Note that we only present the bosonic fields here and the fermionic fields can be read from (\ref{ConformalCompositeVector}) given the single multiplet truncation condition (\ref{SingleMultipletTruncation}). This composite vector multiplet can be used in the two vector multiplet action (\ref{VVprime}), which takes the following form after gauge fixing (\ref{GaugeFixing3})
		\bea
		e^{-1} \cL_{V V^\prime} &=& -\frac14  F^{ab} F_{ab}^\prime - \frac1{16} \e^{abcdef} B_{ab} F_{cd} F_{ef}^\prime +Y^{ij} Y_{ij}^\prime \,,
		\eea 
		and the resulting off-diagonal action, which we refer to as $RY_{ij}$ action, is given by
		\bea
		e^{-1} \cL_{R Y_{ij}} &=& - L^{-1}Y_{ij} D^a D_a L^{ij} + \frac12 L^{-1} Y_{ij} L^{ij}R - \frac1{24} L^{-1} L^{ij} Y_{ij} H_{abc}H^{abc}\nn\\
		&& + L^{-3} Y_{ij} L_{kl} D_a L^{ik} D^a L^{jl} + L^{-3} E^a Y_{ij} L_k{}^{i} D_a L^{jl} + \frac14 L^{-3} Y_{ij} L^{ij} E^a E_a \,,\nn\\
		&& + \frac12 F^{ab} \left(\partial_{a} \left( L^{-1} E_{b} - 2 L^{-1} L_{ij} \cV_{b}{}^{ij} \right) + L^{-3} L_{j}{}^k \partial_{a} L^{jl} \partial_{b} L_{lk} \right) \nn\\
		&& + \frac18  \e^{abcdef} B_{ab} F_{cd}\left(\partial_{e} \left( L^{-1} E_{f} - 2 L^{-1} L_{ij} \cV_{f}{}^{ij} \right) + L^{-3} L_{j}{}^k \partial_{e} L^{jl} \partial_{f} L_{lk} \right)  \,.
		\label{RYAction}
		\eea
		With this off-diagonal action in hand we can obtain an $R^2$ extended Einstein-Maxwell supergravity by considering the following action
		\bea
		\cL = \cL_{EH} + \cL_{V} + \cL_{R Y_{ij}} + g \cL_{VL} \,,
		\label{R2SalamSezgin}
		\eea
		where $\cL_{EH}$ refers to the off-shell Poincar\'e supergravity in Jordan frame (\ref{EHinJF2}), $\cL_{VL}$ refers to the vector-linear coupling (\ref{LocalVL}), $\cL_V$ refers to the off-shell vector multiplet action
		\bea
		e^{-1} \cL_{V} &=& -\frac14  F^{ab} F_{ab} - \frac1{16} \e^{abcdef} B_{ab} F_{cd} F_{ef} +Y^{ij} Y_{ij} \,,
		\label{VectorAction}
		\eea 
		and $\cL_{R Y_{ij}}$ is the off-diagonal $RY_{ij}$ action (\ref{RYAction}). Upon imposing the field equation for $Y_{ij}$, which solves $Y_{ij}$ as
		\bea
		Y^{ij} &=& \frac12 L^{-1} D^a D_a L^{ij} - \frac14 L^{-1} L^{ij}R + \frac1{48} L^{-1} L^{ij} H_{abc}H^{abc} - \frac12 L^{-3} L_{kl} D_a L^{ik} D^a L^{jl}\nn\\
		&&-\frac12  L^{-3} E^a L_k{}^{(i} D_a L^{j)l} - \frac18 L^{-3} L^{ij} E^a E_a - \frac12 g L^{ij} \,,
		\eea
		we obtain $R^2$ action via $Y^{ij} Y_{ij}$ term in the vector multiplet action. As shown in \cite{BergshoeffHD}, an off-shell version of the Salam - Sezgin model \cite{SalamSezgin} can be obtained by a combination of a single linear multiplet truncation of the off-shell Poincar\'e supergravity in Jordan frame (\ref{EHinJF2}), the off-shell vector multiplet action (\ref{VectorAction}) and a local vector-linear multiplet action (\ref{LocalVL}). With the off-diagonal $RY_{ij}$ action, one may improve off-shell Salam - Sezgin model of \cite{BergshoeffHD} with $RY_{ij}$ action, in which case the elimination of the auxiliary field $Y_{ij}$ would lead to an $R^2$ extension of Salam - Sezgin model.
		
		\section{Discussion}\label{Sec5}
		
		In this paper, we provide a systematic analysis of linear multiplets of six dimensional $\cN = (1,0)$ supergravity, which has been shown to be crucial in the construction of higher curvature models \cite{OzkanThesis,6DHD}. Our analysis start with an investigation of rigid linear multiplets, in which case the couplings of linear multiplets are determined by a function $\cF_{IJ}(L)$ that is subject to two mild constraints (\ref{RigidConstraints}). After establishing the relation between the five dimensional $\cN = 2$ and six dimensional $\cN=(1,0)$ rigid linear multiplets, we repeat our analysis for the case of full $\cN=(1,0)$ superconformal symmetry which paves the way for the local supersymmetric couplings of linear multiplets. For the local superconformal models, we work in a dilaton Weyl multiplet background and use superconformal tensor calculus to provide a superconformal linear multiplet action for $n$-number of linear multiplets. For this case, the function $\cF_{IJ}(L)$ picks up two more constraints which are imposed due to dilatation and S-supersymmetry invariance (\ref{ConformalF}). These constraints are also mild and we provide various examples of $\cF_{IJ}$ for the superconformal scenario. Finally, we discussed various gauge fixing procedures and off-shell supergravity models. In particular, we discussed an off-diagonal invariant, which we refer to as the $RY_{ij}$ invariant (\ref{RYAction}), which leads to an $R^2$-extended Einstein-Maxwell supergravity upon imposing the field equations. 
		
		There are various directions to pursue following our work. First, it would be interesting to investigate the supersymmetry solutions of an $R^2$-extended Einstein-Maxwell supergravity. The off-shell model that we have in mind here is given by the equation (\ref{R2SalamSezgin}). Note that there is also an off-shell $R^2$ model constructed in \cite{OzkanThesis,6DHD} which can be added to the combination that we discussed in (\ref{R2SalamSezgin}). It remains to be checked whether these two different paths to improve the Salam-Sezgin model with an $R^2$ term lead to the same physical theory. Second, noticing that a composite superconformal linear multiplet contain a constrained vector $E_a^\prime$ given by (\ref{CompositeLocalLinear2})
		\bea
		E_a^\prime &=& \cD^b (\s \widehat F_{ba}) + \ldots \,,
		\label{CompositeEa}
		\eea
		where we remind the reader that $E_{\m\n}$ is related to $E_a$ via $E^a = e_\m{}^a \cD_\n E^{\m\n}$. Therefore, a linear multiplet action that contains an $E^a E_a$ term would produce a higher derivative vector multiplet action with an $F \Box F$ term. Such a model can easily be obtained with a non-symmetric choice of $\cF_{IJ}$ given in (\ref{NonSymmetricFIJ}) where the unprimed multiplet can be used as a compensating multiplet and the primed multiplet can be used as a composite linear multiplet. Note that as $\cF_{22} = L^{-1}$ for the non-symmetric choice, an action for such an $\cF_{IJ}$ would contain a term $L^{-1} E_a^\prime E^{\prime a}$ due to $\cF_{IJ} E_a^I E^{a J}$ in (\ref{LocalLinear}). Thus, upon using the composite expression for $E_a$ in (\ref{CompositeEa}) one would obtain the desired higher derivative vector multiplet action. This result should be compared with \cite{FBoxF1,FBoxF2}. Finally, it would be interesting to see if other off-diagonal invariants can be constructed and if they can have interesting physical implications in higher derivative extended supersymmetric models.
		
		\section*{Acknowledgment}
		The work of U.A is supported by TUBITAK grant 118F091. M.O. is supported in part by TUBITAK grant 118F091. 
		
		\appendix
		
		\section{Weyl Multiplets of $D=6,\,\cN=(1,0)$ Supergravity}{\label{Appendix1}}
		
		In this section, we briefly review the elements of six dimensional superconformal tensor calculus. We refer \cite{Bergshoeff6D,Coomans6D,CoomansThesis} to readers interested in a more detailed treatment. The six dimensional $\cN = (1,0)$ conformal tensor calculus is based on the exceptional superalgebra $OSp(6,2|1)$ with the generators
		\bea
		P_a\,, M_{ab}\,, D\,, K_a \,, U_{ij} \,, Q_\alpha^i\,, S_\alpha^i \,,
		\eea
		with the corresponding gauge fields
		\bea
		e_\m{}^a\,, \o_\m{}^{ab}\,, b_\m \,, f_\m{}^a\,, \cV_\m{}^{ij} \,, \psi^i_\m \,, \phi_\m^i \,,
		\label{OSpGaugeFields}
		\eea
		where $a,b,\ldots $ are the Lorentz indices $\m,\n\, \ldots$ are the world vector indices, $\a$ is a spinor index and $i,j = 1, 2$ is an $SU(2)$ index. In the set of generators of the superconformal algebra, $\{P_a, M_{ab}, D, K_a \}$ represent the generators of the conformal algebra. For the remaining generators, $U_{ij}$ is the $SU(2)$ generator and $Q_\alpha^i$ and $S_\alpha^i$ are the generators of the $Q-$SUSY and the $S-$SUSY respectively.
		
		A set of constraints, known as the conventional constraints, can be applied to the set of gauge fields (\ref{OSpGaugeFields}), which leaves $\left(e_\m{}^a\,, b_\m\,, \cV_{\m}{}^{ij}, \p_\m^i\right)$ as independent fields and $\left(\o_\m{}^{ab}\,, f_\m{}^a\,, \phi_\m^i\right)$ becomes dependent. However, a simple counting argument shows that the number of bosonic and fermionic degrees of freedom do not match and one needs to include matter fields to form a Weyl multiplet. One possible choice, leading to the standard Weyl multiplet, is the inclusion of a real scalar field $D$, an antisymmetric tensor of negative duality $T_{abc}^-$ and an $SU(2)$ Majorana-Weyl spinor of negative chirality $\chi^i$. The $Q$-SUSY, $S$-SUSY and special conformal transformation rules are given by \cite{Bergshoeff6D,Coomans6D}
		\bea
		\delta e_{\mu}{}^a &=& \frac{1}{2}\bar{\epsilon}\gamma^a\psi_{\mu}\,, \nonumber \\
		\delta \psi_{\mu}^i &=& \partial_{\mu}\epsilon^i+\frac{1}{2}b_{\mu}\epsilon^i+\frac{1}{4}\omega_{\mu}{}^{ab}\gamma_{ab}\epsilon^i+{\cal V}_{\mu}{}^i{}_j\epsilon^j +\frac{1}{24}\gamma\cdot T^-\gamma_{\mu}\epsilon^i +\gamma_{\mu}\eta^i \,,\nn\\
		\delta b_{\mu} &=&-\frac{1}{2}\bar{\epsilon}\phi_{\mu}-\frac{1}{24}\bar{\epsilon}\gamma_{\mu}\chi+\frac{1}{2}\bar{\eta}\psi_{\mu}-2\lambda_K{}_{\mu}, \nonumber \\
		\delta {\cal V}_{\mu}^{ij} &=& 2\bar{\epsilon}^{(i}\phi_{\mu}^{j)}
		+2\bar{\eta}^{(i}\psi_{\mu}^{j)}+\frac{1}{6}\bar{\epsilon}^{(i}\gamma_{\mu}\chi^{j)} \,,\nn\\
		\delta T_{abc}^-&=&-\frac{1}{32}\bar{\epsilon}\gamma^{de}\gamma_{abc} {R}_{de}(Q)-\frac{7}{96}\bar{\epsilon}\gamma_{abc}\chi, \nonumber \\
		\delta \chi^i&=&\frac{1}{8}\bigl(\cD_\mu  \gamma\cdot T^-\bigr)\gamma^{\mu}\epsilon^i -\frac{3}{8}\gamma\cdot{R}^{ij}({\cal V})\epsilon_j+\frac{1}{4}D\epsilon^i+\frac{1}{2}\gamma\cdot T^-\eta^i\,, \nonumber \\
		\delta D &=& \bar{\epsilon}\gamma^{\mu}\cD_{\mu}\chi-2\bar{\eta}\chi\,,
		\eea
		where the covariant curvatures are given by
		\bea
		\mathcal{D}_{\mu}T^-_{abc}&=&\partial_{\mu}T^-_{abc}-3\omega_{\mu}{}^d{}_{[a}T^-_{bc]d}-b_{\mu}T^-_{abc}+\frac{1}{32}\opsi_{\mu}\gamma^{de}\gamma_{abc}{R}_{de}(Q)+\frac{7}{96}\opsi_{\mu}\gamma_{abc}\chi\,,  \nn\\
		\mathcal{D}_{\mu}\chi^i&=&\Bigl(\partial_{\mu}-\frac{3}{2}b_{\mu}+\frac{1}{4}\omega_{\mu}{}^{ab}\gamma_{ab}\Bigr)\chi^i+{\cal V}_{\mu}{}^i{}_j\chi^j-\frac{1}{8}\Bigl(\mathcal{D}_{\nu}\gamma\cdot T^-\Bigr)\gamma^{\nu}\psi_{\mu}^i \nonumber \\
		&&+\frac{3}{8}\gamma\cdot {R}^{ij}({\cal V})\psi_{\mu j}-\frac{1}{4}D\psi_{\mu}^i-\frac{1}{2}\gamma\cdot T^- \phi_{\mu}^i\,,
		\eea
		and the relevant modified group theoretical curvatures are
		\bea
		R_{\m\n}{}^{a} (P) &=& 2\partial_{[\mu}e_{\nu]}{}^a+2b_{[\mu}e_{\nu]}{}^a+2\omega_{[\mu}{}^{ab}e_{\nu]b}-\frac{1}{2}\bar{\psi}_{\mu}\gamma^a\psi_{\nu}\,,\nn\\
		R_{\m\n}{}^{ab} (M) &=& 2\partial_{[\mu}\omega_{\nu]}{}^{ab}+2\omega_{[\mu}{}^{ac}\omega_{\nu]c}{}^b-8f_{[\mu}{}^{[a}e_{\nu]}{}^{b]}+\bar{\psi}_{[\mu}\gamma^{ab}\phi_{\nu]} \nonumber \\
		&&+\bar{\psi}_{[\mu}\gamma^{[a}{R}_{\nu]}{}^{b]}(Q)+\frac{1}{2}\bar{\psi}_{[\mu}\gamma_{\nu]}{R}^{ab}(Q)-\frac{1}{6}e_{[\mu}{}^{[a}\bar{\psi}_{\nu]}\gamma^{b]}\chi-\frac{1}{2}\bar{\psi}_{\mu}\gamma_c\psi_{\nu}T^-{}^{abc}\,,\nonumber \\ 
		R_{\m\n}{}^{ij} (\cV) &=& 2\partial_{[\mu}{\cal V}_{\nu]}{}^{ij}-2{\cal V}_{[\mu}{}^{k(i}{\cal V}_{\nu]}{}^{j)}{}_k
		-4\bar{\psi}_{[\mu}{}^{(i}\phi_{\nu]}{}^{j)}-\frac{1}{3}\bar{\psi}_{[\mu}{}^{(i}\gamma_{\nu]}\chi^{j)} \,,\nn\\
		R_{\m\n}{}^i (Q) &=& 2\Bigl(\partial_{[\mu}+\frac{1}{2}b_{[\mu}+\frac{1}{4}\omega_{[\mu}{}^{ab}\gamma_{ab}\Bigr)\psi_{\nu]}^i
		+2{\cal V}_{[\mu}{}^i{}_j\psi_{\nu]}^j-\frac{1}{12}\gamma \cdot T^- \gamma_{[\nu}\psi_{\mu]}^i-2\gamma_{[\nu}\phi_{\mu]}^i  \,.
		\eea
		Within the standard Weyl multiplet, the dependent fields are given by
		\bea
		\omega_{\mu}{}^{ab}&=&2e^{\nu [a}\partial_{[\mu}e_{\nu]}{}^{b]}-e^{\rho[a}e^{b]\sigma}e_{\mu}{}^{c}\partial_{\rho}e_{\sigma c}+\frac{1}{4}\bigl(2\bar{\psi}_{\mu}\gamma^{[a}\psi^{b]}+\bar{\psi}^a\gamma_{\mu}\psi^b\bigr)+2e_{\mu}{}^{[a}b^{b]}\,,  \nonumber \\
		f_{\mu}{}^a&=&\frac{1}{8}\bigl({R}'_{\mu}{}^a(M)-\frac{1}{10}e_{\mu}{}^a {R}'(M)\bigr)-\frac{1}{8}T_{\mu cd}^-T^{-}{}^{acd}+\frac{1}{240}e_{\mu}{}^aD\,, \nonumber \\
		\phi_{\mu}^i&=&-\frac{1}{16}\bigl(\gamma^{ab}\gamma_{\mu}-\frac{3}{5}\gamma_{\mu}\gamma^{ab}\bigr){R}'_{ab}{}^i(Q)-\frac{1}{60}\gamma_{\mu}\chi^i\,, 
		\eea
		Here, the primed quantities are defined as
		\bea
		R^\prime_{\m\n}{}^{ab}(M) \equiv R_{\m\n}{}^{ab}(M) + 8f_{[\mu}{}^{[a}e_{\nu]}{}^{b]}\,, \qquad R^\prime_{\m\n}{}^i (Q) =R^{\m\n}{}^i (Q) + 2\gamma_{[\nu}\phi_{\mu]}^i  \,. 
		\eea
		The definitions of the dependent fields are equivalent to imposing the following set of constraints on the group theoretical curvatures
		\bea
		0 &=& R_{\m\n}{}^a (P) \,,\nn\\
		0 &=& e^\nu{}_b R_{\m\n}{}^{ab} (M) - T_{\m b c}^- T^{-abc} + \frac{1}{12} e_\m{}^a D \,,\nn\\
		0 &=& \g^\m R_{\m\n} (Q) + \frac16 \g_\n \chi^i \,.
		\eea
		Alternatively, one may consider to add a dilaton field $\s$, a dilatino field $\p^i$ and a two-form gauge field $B_{\m\n}$ to the content of gauge fields to match the bosonic and fermionic degrees of freedom. This is most straightforwardly obtained by coupling the standard Weyl multiplet to a tensor multiplet which consists of  a dilaton field $\s$, a dilatino field $\p^i$ and a self-dual antisymmetric tensor field $F_{abc}^+$. The closure of the superconformal algebra on these fields imposes the following constraints \cite{Bergshoeff6D,Coomans6D}
		\bea
		0 &=& \slashed\cD \p^i -\frac16 \s \chi^i - \frac{1}{12} \g \cdot T^- \p^i \,,\nn\\
		0 &=& \Box^c \s - \frac{1}{6} D \s + \frac13 F^+ \cdot T^- + \frac76 \bar\chi \p \,,\nn\\
		0 &=& \cD^c \left( F_{abc}^+ - 2 \s T_{abc}^- \right) - \bar{R}_{ab} (Q) \p - \frac16 \bar\chi \g_{ab} \p \,,
		\eea
		where the relevant covariant curvatures are defined as
		\bea
		\cD_\m \p^i &=& \bigl(\partial_{\mu}-\frac{5}{2}b_{\mu}+\frac{1}{4}\omega_{\mu}{}^{ab}\gamma_{ab}\bigr)\psi^i+{\cal V}_{\mu}{}^i{}_j\psi^j-\frac{1}{48}\gamma\cdot F^+\psi_{\mu}^i-\frac{1}{4}\slashed{\mathcal{D}}\sigma\psi_{\mu}^i+\sigma\phi_{\mu}^i \,,\nn\\
		\cD_{\mu}\sigma&=&\bigr(\partial_{\mu}-2b_{\mu}\bigl)\sigma-\bar{\psi}_{\mu}\psi\,.
		\eea
		These constraints can be solved to relate the fields of standard Weyl multiplet to the dilaton Weyl multiplet. As we mostly worked out the bosonic part of supersymmetric actions, here we will present the bosonic part of the map between the multiplets and the full supersymmetric map can be found in \cite{Coomans6D} in our conventions
		\bea
		D &=& \frac{15}{4} \s^{-1} \Box \s - \frac34 R + \frac5{16} H_{\m\n\r} H^{\m\n\r} \,,\nn\\
		F^+_{\m\n\r} + 2 \s T_{\m\n\r}^- &=& H_{\m\n\r} \,,
		\eea
		where $R$ is the Ricci scalar. Finally, the $Q$-SUSY $S$-SUSY and special conformal transformation rules for the dilaton Weyl multiplet are
		\bea
		\delta e_{\mu}{}^a &=& \frac{1}{2}\bar{\epsilon}\gamma^a\psi_{\mu}\,, \nonumber \\
		\delta \psi_{\mu}^i &=& \partial_{\mu}\epsilon^i+\frac{1}{2}b_{\mu}\epsilon^i+\frac{1}{4}\omega_{\mu}{}^{ab}\gamma_{ab}\epsilon^i+{\cal V}_{\mu}{}^i{}_j\epsilon^j +\frac{1}{48} \s^{-1} \gamma\cdot H \gamma_{\mu}\epsilon^i +\gamma_{\mu}\eta^i \,,\nn\\
		\delta b_{\mu} &=&-\frac{1}{2}\bar{\epsilon}\phi_{\mu}-\frac{1}{24}\bar{\epsilon}\gamma_{\mu}\chi+\frac{1}{2}\bar{\eta}\psi_{\mu}-2\lambda_K{}_{\mu}, \nonumber \\
		\delta {\cal V}_{\mu}^{ij} &=& 2\bar{\epsilon}^{(i}\phi_{\mu}^{j)}
		+2\bar{\eta}^{(i}\psi_{\mu}^{j)}+\frac{1}{6}\bar{\epsilon}^{(i}\gamma_{\mu}\chi^{j)} \,,\nn\\
		\d \s &=& \bar\e \p  \,,\nn\\
		\d B_{\m\n} &=& - \bar\e \g_{\m\n} \p - \s \bar\e \g_{[\m} \p_{\n]} \,,\nn\\
		\d \p^i &=& \frac{1}{48} \g \cdot H \e^i + \frac14 \slashed\cD \s \e^i - \s \eta^i \,.
		\eea
		
		\newpage
		\providecommand{\href}[2]{#2}\begingroup\raggedright\endgroup
		\newpage


\begin{thebibliography}{99}
			
			\bibitem{6DRef1}
			H.~Nishino and E.~Sezgin,
			``Matter and Gauge Couplings of N=2 Supergravity in Six-Dimensions,''
			Phys. Lett. B \textbf{144} (1984), 187-192
			
			\bibitem{6DRef2}
			H.~Nishino and E.~Sezgin,
			``The Complete $N=2$, $d=6$ Supergravity With Matter and \{Yang-Mills\} Couplings,''
			Nucl. Phys. B \textbf{278} (1986), 353-379
			
			\bibitem{6DRef3}
			H.~Nishino and E.~Sezgin,
			``New couplings of six-dimensional supergravity,''
			Nucl. Phys. B \textbf{505} (1997), 497-516
			
			\bibitem{6DRef4}
			A.~Sagnotti,
			``A Note on the Green-Schwarz mechanism in open string theories,''
			Phys. Lett. B \textbf{294} (1992), 196-203
			
			\bibitem{6DRef5}
			S.~Ferrara, R.~Minasian and A.~Sagnotti,
			``Low-energy analysis of M and F theories on Calabi-Yau threefolds,''
			Nucl. Phys. B \textbf{474} (1996), 323-342
			
			\bibitem{6DRef6}
			S.~Ferrara, F.~Riccioni and A.~Sagnotti,
			``Tensor and vector multiplets in six-dimensional supergravity,''
			Nucl. Phys. B \textbf{519} (1998), 115-140
			
			\bibitem{6DRef7}
			F.~Riccioni,
			``All couplings of minimal six-dimensional supergravity,''
			Nucl. Phys. B \textbf{605} (2001), 245-265
			
			\bibitem{SalamSezgin}
			A.~Salam and E.~Sezgin,
			``Chiral Compactification on Minkowski x S**2 of N=2 Einstein-Maxwell Supergravity in Six-Dimensions,''
			Phys. Lett. B \textbf{147} (1984), 47
			
			\bibitem{SalamSezginString}
			M.~Cvetic, G.~W.~Gibbons and C.~N.~Pope,
			``A String and M theory origin for the Salam-Sezgin model,''
			Nucl. Phys. B \textbf{677} (2004), 164-180
			
			\bibitem{MinasianLiu1}
			J.~T.~Liu and R.~Minasian,
			``Higher-derivative couplings in string theory: dualities and the B-field,''
			Nucl. Phys. B \textbf{874} (2013), 413-470	
			
			\bibitem{6DGB}
			J.~Novak, M.~Ozkan, Y.~Pang and G.~Tartaglino-Mazzucchelli,
			``Gauss-Bonnet supergravity in six dimensions,''
			Phys. Rev. Lett. \textbf{119} (2017) no.11, 111602
			
			\bibitem{MinasianLiu2}
			J.~T.~Liu and R.~Minasian,
			``Higher-derivative couplings in string theory: five-point contact terms,''
			[arXiv:1912.10974 [hep-th]].
			
			
			
			\bibitem{deBoer}
			J.~de Boer,
			``Six-dimensional supergravity on S**3 x AdS(3) and 2-D conformal field theory,''
			Nucl. Phys. B \textbf{548} (1999), 139-166
			
			\bibitem{Bergshoeff6D}
			E.~Bergshoeff, E.~Sezgin and A.~Van Proeyen,
			``Superconformal Tensor Calculus and Matter Couplings in Six-dimensions,''
			Nucl. Phys. B \textbf{264} (1986), 653
			
			
			\bibitem{SCTC1}
			M.~Kaku, P.~K.~Townsend and P.~van Nieuwenhuizen,
			``Properties of Conformal Supergravity,''
			Phys. Rev. D \textbf{17} (1978), 3179
			
			\bibitem{SCTC2}
			M.~Kaku and P.~K.~Townsend,
			``POINCARE SUPERGRAVITY AS BROKEN SUPERCONFORMAL GRAVITY,''
			Phys. Lett. B \textbf{76} (1978), 54-58
			
			\bibitem{SCTC3}
			S.~Ferrara, M.~Kaku, P.~K.~Townsend and P.~van Nieuwenhuizen,
			``Gauging the Graded Conformal Group with Unitary Internal Symmetries,''
			Nucl. Phys. B \textbf{129} (1977), 125-134
			
			\bibitem{SCTC4}
			M.~Kaku, P.~K.~Townsend and P.~van Nieuwenhuizen,
			``Gauge Theory of the Conformal and Superconformal Group,''
			Phys. Lett. B \textbf{69} (1977), 304-308
			
			\bibitem{6DHD}
			D.~Butter, J.~Novak, M.~Ozkan, Y.~Pang and G.~Tartaglino-Mazzucchelli,
			``Curvature squared invariants in six-dimensional ${\cal N} = (1,0)$ supergravity,''
			JHEP \textbf{04} (2019), 013
			
			\bibitem{Sokatchev}
E.~Sokatchev,
``Off-shell Six-dimensional Supergravity in Harmonic Superspace,''
Class. Quant. Grav. \textbf{5} (1988), 1459-1471
			
			\bibitem{O2nMultiplet}
W.~D.~Linch, III and G.~Tartaglino-Mazzucchelli,
``Six-dimensional Supergravity and Projective Superfields,''
JHEP \textbf{08} (2012), 075
			
			\bibitem{OzkanThesis}
			M.~Ozkan,
			``Supersymmetric Curvature Squared Invariants in Five and Six Dimensions,''
			
			
			
			\bibitem{deWit4D}
			B.~de Wit and F.~Saueressig,
			``Off-shell N=2 tensor supermultiplets,''
			JHEP \textbf{09} (2006), 062
			
			\bibitem{Ozkan5D}
			M.~Ozkan,
			``Off-shell $ \mathcal{N} $ = 2 linear multiplets in five dimensions,''
			JHEP \textbf{11} (2016), 157
			
			\bibitem{Coomans6D}
			F.~Coomans and A.~Van Proeyen,
			``Off-shell N=(1,0), D=6 supergravity from superconformal methods,''
			JHEP \textbf{02} (2011), 049
			[erratum: JHEP \textbf{01} (2012), 119]
			
			\bibitem{Ozkan1}
			F.~Coomans and M.~Ozkan,
			``An off-shell formulation for internally gauged D=5, N=2 supergravity from superconformal methods,''
			JHEP \textbf{01} (2013), 099
			
			\bibitem{Bergshoeff5D1}
			E.~Bergshoeff, S.~Cucu, T.~De Wit, J.~Gheerardyn, R.~Halbersma, S.~Vandoren and A.~Van Proeyen,
			``Superconformal N=2, D = 5 matter with and without actions,''
			JHEP \textbf{10}, 045 (2002)
			
			\bibitem{Bergshoeff5D2}
			E.~Bergshoeff, S.~Cucu, T.~de Wit, J.~Gheerardyn, S.~Vandoren and A.~Van Proeyen,
			``N = 2 supergravity in five-dimensions revisited,''
			Class. Quant. Grav. \textbf{21} (2004), 3015-3042
			
			\bibitem{BergshoeffSezgin5D}
			E.~A.~Bergshoeff, J.~Rosseel and E.~Sezgin,
			``Off-shell D=5, N=2 Riemann Squared Supergravity,''
			Class. Quant. Grav. \textbf{28} (2011), 225016
			
			\bibitem{BergshoeffHD}
			E.~Bergshoeff, F.~Coomans, E.~Sezgin and A.~Van Proeyen,
			``Higher Derivative Extension of 6D Chiral Gauged Supergravity,''
			JHEP \textbf{07} (2012), 011	
			
			\bibitem{Cremonini5D}
S.~Cremonini, K.~Hanaki, J.~T.~Liu and P.~Szepietowski,
``Black holes in five-dimensional gauged supergravity with higher derivatives,''
JHEP \textbf{12} (2009), 045
			
			\bibitem{SuperBHT}
			E.~A.~Bergshoeff, O.~Hohm, J.~Rosseel, E.~Sezgin and P.~K.~Townsend,
			``More on Massive 3D Supergravity,''
			Class. Quant. Grav. \textbf{28} (2011), 015002
			
			\bibitem{BISugra}
			E.~Bergshoeff and M.~Ozkan,
			``3D Born-Infeld Gravity and Supersymmetry,''
			JHEP \textbf{08} (2014), 149
			
			\bibitem{3DN2}
			G.~Alka\c{c}, L.~Basanisi, E.~A.~Bergshoeff, M.~Ozkan and E.~Sezgin,
			``Massive $ \mathcal{N} $ = 2 supergravity in three dimensions,''
			JHEP \textbf{02} (2015), 125
			
			
			
			
			
			\bibitem{CoomansThesis}
			F.~Coomans,
			``Higher derivative supergravities from superconformal methods,''
			
			\bibitem{FBoxF1}
			D.~Butter, S.~M.~Kuzenko, J.~Novak and S.~Theisen,
			``Invariants for minimal conformal supergravity in six dimensions,''
			JHEP \textbf{12} (2016), 072
			
			\bibitem{FBoxF2}
			D.~Butter, J.~Novak and G.~Tartaglino-Mazzucchelli,
			``The component structure of conformal supergravity invariants in six dimensions,''
			JHEP \textbf{05} (2017), 133
			
		\end{thebibliography}
	\end{document}